\documentclass[epj]{webofc}
\usepackage[utf8]{inputenc}
\usepackage[varg]{txfonts}   % Web of Conferences font
\usepackage{booktabs}
\usepackage{xcolor}
\usepackage{tabu}
\definecolor{darkred}{rgb}{0.4,0.0,0.0}
\definecolor{darkgreen}{rgb}{0.0,0.4,0.0}
\definecolor{darkblue}{rgb}{0.0,0.0,0.4}
\usepackage[bookmarks,linktocpage,colorlinks,
    linkcolor = darkred,
    urlcolor  = darkblue,
    citecolor = darkgreen]{hyperref}
\usepackage{subfigure}
\usepackage{caption}
\wocname{EPJ Web of Conferences}
\woctitle{Lattice2017}
\begin{document}
\selectlanguage{english}
\title{\boldmath $\overline{B}\rightarrow D^\ast\ell\overline{\nu}$ at non-zero recoil}

\author
{ \firstname{Alejandro} \lastname{Vaquero Avilés-Casco}\inst{1}\fnsep\thanks{Speaker, \email{alexvaq@physics.utah.edu}} \and
  \firstname{Carleton}  \lastname{DeTar}\inst{1}	\and
  \firstname{Daping}    \lastname{Du}\inst{2}		\and
  \firstname{Aida}      \lastname{El-Khadra}\inst{3,4}	\and
  \firstname{Andreas}   \lastname{Kronfeld}\inst{4,5}	\and
  \firstname{John}      \lastname{Laiho}\inst{2}	\and
  \firstname{Ruth S.}   \lastname{Van de Water}\inst{4}	}

\institute
{ Department of Physics and Astronomy, University of Utah, Salt Lake City, Utah, 84112,  USA \and
  Department of Physics, Syracuse University, Syracuse, New York, 13244, USA		     \and
  Department of Physics, University of Illinois, Urbana, Illinois, 61801, USA		     \and
  Fermi National Accelerator Laboratory, Batavia, Illinois, 60510, USA			     \and
  Institute for Advanced Study, Technische Universität München, 85748 Garching, Germany }

\abstract 
{ We present preliminary results from our analysis of the form factors for the $\overline{B}\rightarrow D^\ast\ell\overline{\nu}$ decay at non-zero recoil. Our analysis includes 15 MILC
asqtad ensembles with $N_f=2+1$ flavors of sea quarks and lattice spacings ranging from $a\approx 0.15$ fm down to $0.045$ fm. The valence light quarks employ the asqtad action, whereas the
heavy quarks are treated using the Fermilab action. We conclude with a discussion of future plans and phenomenological implications. When combined with experimental measurements of the
decay rate, our calculation will enable a determination of the CKM matrix element $\left|V_{cb}\right|$. }

\maketitle

\section{Introduction}\label{intro}
Although the Standard Model (SM) is widely regarded as a very successful theory, its description of nature is not fully satisfactory, and it is believed to be incomplete. This fact has
set a large part of the scientific community in the search for physics Beyond the Standard Model (BSM). One of the most promising places to look for new physics is the flavor sector of
the SM. In particular, the unitarity triangle and the CKM matrix elements \cite{Cabibbo:1963yz,Kobayashi:1973fv} have been object of extensive research \cite{Antonelli:2009ws}, as newer
methods and algorithms improved the precision of the measurements. The latest experimental averages \cite{Patrignani:2016xqp,Amhis:2016xyh} reveal a tension of about $3\sigma$ between the
inclusive and exclusive determinations of $\left|V_{cb}\right|$, and it is of foremost importance to investigate the origins of this tension.

One way to obtain $\left|V_{cb}\right|$ is from the decay rate of $\overline{B}\rightarrow D^\ast\ell\overline{\nu}$. Experimental measurements of this decay rate are precise for large
momenta, but noisy at low recoil, due to a supression of the phase space \cite{Patrignani:2016xqp}. The lattice approach, on the other hand, excels at low recoil, where it becomes more
precise. Forthcoming experiments at LHCb \cite{Aaij:2244311} and Belle-II \cite{Abe:2010gxa}, are expected to reduce the experimental uncertainties, so improved lattice calculations are
needed to maximize their impact on $\left|V_{cb}\right|$ determinations. In this work we present first, preliminary results of our lattice QCD calculation of the form factors for the
$\overline{B}\rightarrow D^\ast\ell\overline{\nu}$ process at non-zero recoil.

\section{Form factor definitions and ratios}\label{ffDef}
The $\overline{B}\rightarrow D^\ast\ell\overline{\nu}$ decay is mediated by a $V-A$ weak current. The matrix elements can be decomposed in the following way:
\begin{align}
\frac{\left\langle D^\ast(p_{D^\ast},\epsilon^\nu)\right|\mathcal{V}^\mu\left|\bar{B}(p_B)\right\rangle}{2\sqrt{M_B\,M_{D^\ast}}} &= 
\frac{1}{2}\epsilon^{\nu *}\varepsilon^{\mu\nu}_{\,\,\rho\sigma} v_B^\rho v_{D^\ast}^\sigma h_V(w),\label{vector} \\
\frac{\left\langle D^\ast(p_{D^\ast},\epsilon^\nu)\right|\mathcal{A}^\mu\left|\bar{B}(p_B)\right\rangle}{2\sqrt{M_B\,M_{D^\ast}}} &= 
\frac{i}{2}\epsilon^{\nu *}\left[g^{\mu\nu}\left(1+w\right)h_{A_1}(w) - v_B^\nu\left(v_B^\mu h_{A_2}(w) + v_{D^\ast}^\mu h_{A_3}(w)\right)\right],\label{axial}
\end{align}
where $\mathcal{V}^\mu$ and $\mathcal{A}^\mu$ are the vector and axial, continuum, electroweak currents, respectively. The 4-vectors $v_X^\mu\quad (X=\overline{B},D^\ast)$ are the
4-velocities, defined as $v_X^\mu = p_X^\mu/M_X$, $w = v_B\cdot v_{D^\ast}$ is the recoil parameter, the $\epsilon^\nu$ is the polarization of the $D^\ast$ final stat eand the
$h_k(w)\quad (k=V,A_{1,2,3})$ are the different form factors that contribute to the decay rate.

The differential decay rate can be expressed as
\begin{equation}
\frac{d\Gamma}{dw} = \frac{G_F^2 M_B^5}{4\pi^3}r^3(1-r)^2(w^2-1)^{\frac{1}{2}}\left|\eta_{EW}\right|^2\left|V_{cb}\right|^2\chi(w)\left|\mathcal{F}(w)\right|^2,
\end{equation}
with $r = M_{D^\ast}/M_B$, the ratio of the meson rest masses and $\eta_{EW}$, the electroweak correction factor for NLO box diagrams. The kinematic factor $\sqrt{w^2-1}$
is what makes experimental measurements so difficult at low recoil. The functions $\chi(w)$ and $\mathcal{F}(w)$ are standard, motivated by the heavy-quark limit:

\begin{equation}
    \chi(w) = \left(1+w\right)^2\lambda(w)\qquad{\rm with }\qquad \lambda(w) = \frac{1 + \frac{4w}{w+1}t^2(w)}{12},\qquad t^2(w) = \frac{1-2wr+r^2}{(1-r)^2},
\end{equation}
\begin{equation}
    \mathcal{F}(w) = h_{A_1}(w)\sqrt{\frac{H_0^2(w) + H_+^2(w) + H_-^2(w)}{\lambda(w)}},
\end{equation}

\noindent where

\begin{minipage}{0.481\textwidth}
  \begin{equation}
    X_V(w) = \sqrt{\frac{w-1}{w+1}}\frac{h_V(w)}{h_{A_1}(w)},\label{XV}
  \end{equation}
\end{minipage}%
\begin{minipage}{0.481\textwidth}
  \begin{equation}
    X_2(w) = (w-1)\frac{h_{A_2}(w)}{h_{A_1}(w)},\label{X2}
  \end{equation}
\end{minipage}

\begin{minipage}{0.481\textwidth}
  \begin{equation}
    X_3(w) = (w-1)\frac{h_{A_3}(w)}{h_{A_1}(w)},\label{X3}
  \end{equation}
\end{minipage}%
\begin{minipage}{0.481\textwidth}
  \begin{equation}
    H_0(w) = \frac{w - r - X_3(w) - rX_2(w)}{1-r},
  \end{equation}
\end{minipage}

\begin{equation}
  H_\pm(w) = t(w)\left(1\mp X_V(w)\right),
\end{equation}

\noindent The calculation of $\mathcal{F}(w)$ will allow us to determine $\left|V_{cb}\right|$ from experimental measurements of the differential decay rate.

\section{Simulation details}\label{data}

\begin{figure}[h]
  \centering
  \sidecaption
  \includegraphics[width=0.38\linewidth,angle=0]{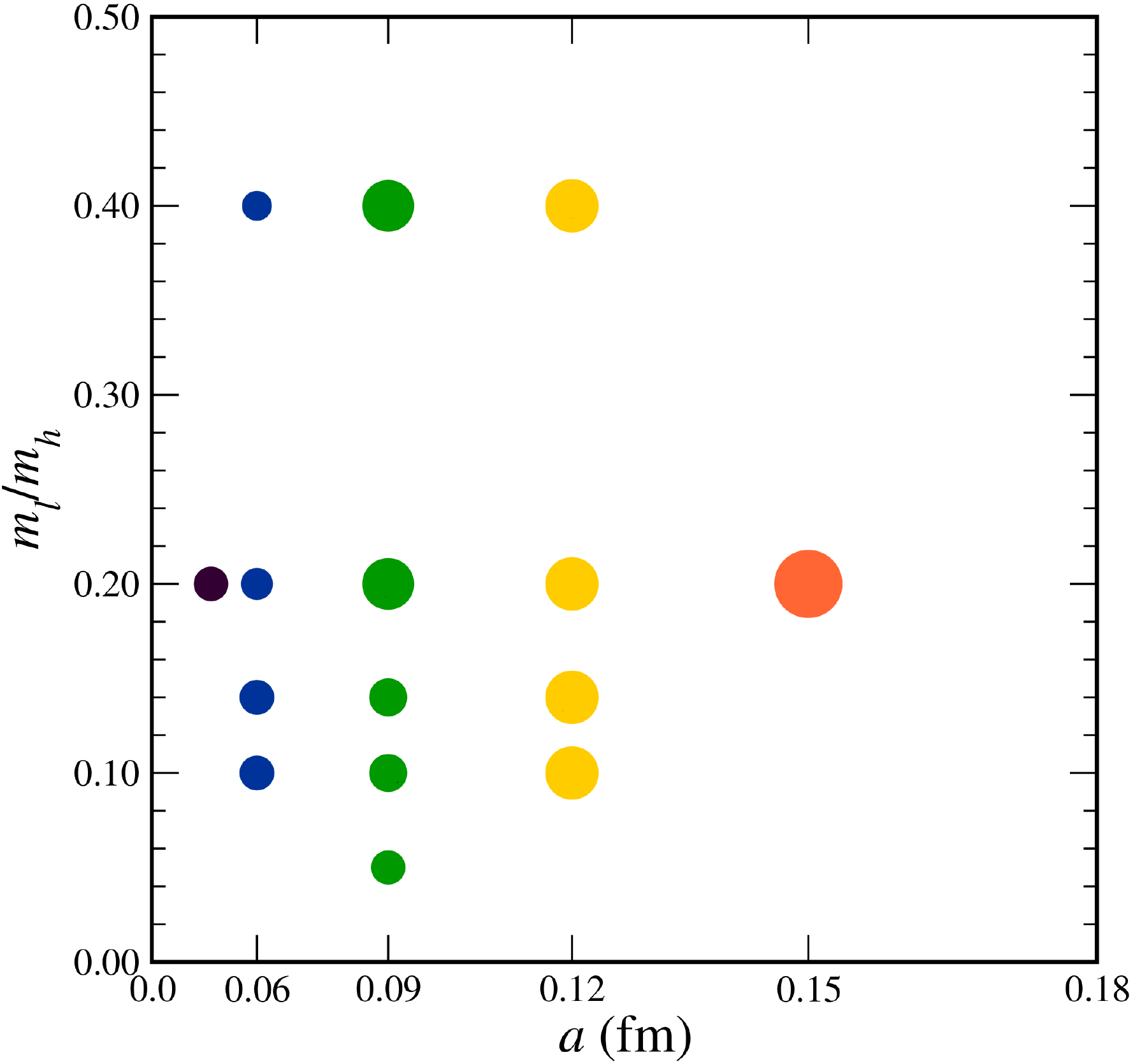}
  \caption{List of available ensembles. The area of the circle is proportional to the statistics. \label{ensList}}
\end{figure}

For this analysis we use the same set of ensembles as in the zero recoil publication \cite{Bailey:2014tva}. These are 15 different MILC asqtad ensembles with $N_f=2+1$ flavors of sea
quarks \cite{Bazavov:2009bb}. The valence light quarks are simulated with the asqtad action \cite{Lepage:1998vj}, whereas the heavy quarks use the clover action with the Fermilab
interpretation \cite{ElKhadra:1996mp}. The calculations on these ensembles are performed at ${\bf p}^2 = 0, \left(\frac{2\pi}{L}\right)^2, \left(\frac{4\pi}{L}\right)^2$ in lattice units,
with $L$ the spatial size of the lattice. We distinguish two different cases: momentum perpendicular to the $D^\ast$ polarization and an average over parallel and perpendicular directions. 
Fig.~\ref{ensList} shows the size of the ensemble as a function of the lattice spacing and the quark masses. For further details we refer to tables I and II of \cite{Bailey:2014tva}.

\section{Calculation of correlation functions}
\subsection{Two-point functions and calculation of rest masses and excited states}

The 2-point functions give us information about the energy levels of the states, as well as the overlap factors $Z$. In this work, the definitions and notation of the 2-point functions
follow ref.~\cite{Lattice:2015rga}. We fit 2-point functions for the $\bar{B}$ and the $D^\ast$ meson with smeared $(1S)$ and point $(d)$ sources at source and sink, giving rise to four
possible combinations per ensemble and momentum: $(1S,1S)$, $(d,d)$, $(1S,d)$ and $(d,1S)$, where we combined the latter two into a symmetric average.

For our analysis, we perform the joint fit of the three 2-point functions with different smearings at the same time on each ensemble. At non-zero momentum ($D^\ast$), we also distinguished
between momentum parallel or perpendicular with respect to the meson polarization. Therefore at non-zero momentum our simultaneous fits involve six 2-point functions. We use Bayesian
constraints with loosely constrained priors. The main function of the priors here was to prevent the fitter from arriving at absurd values. The prior errors were large enough so small
variations in their values didn't change the final results. The correlation matrix for the fitter was obtained using the jackknife method.

We vary the number of states in our fit function: $1+1$, $2+2$ and $3+3$, where the first figure is the number of standard states, and the second gives the number of oscillating states.
We choose the $2+2$ fits, which balance reasonable $p$~values results with low errors, and are consistent with the results coming from the $3+3$ fits. The initial time of the fit is chosen
after reaching a plateau in the value of the meson mass, and it is adjusted to be similar in physical units for all the ensembles. The final time is taken to ensure the resulting $p$~value
is reasonable, i.e., we check that the $p$~values follow a uniform distribution.

\begin{figure}[h]
  \centering
  \sidecaption
  \includegraphics[width=0.60\linewidth,angle=0]{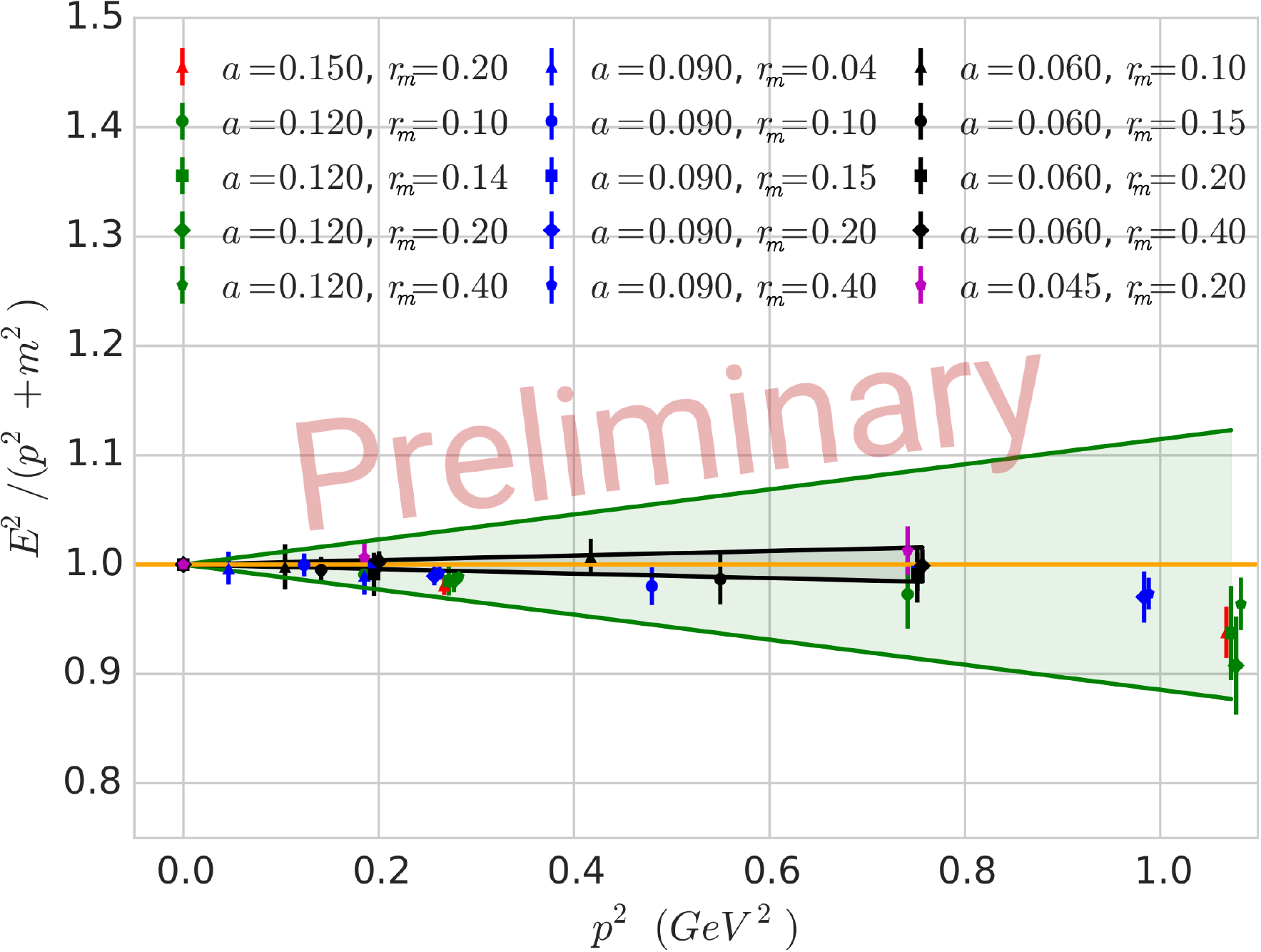}
  \caption{Dispersion relation for the $D^\ast$ meson, all ensembles. The legend gives the lattice spacing $a$ and the ratio between the light and the strange quark masses, called here
	   $r_m$. The cones represent the expected discretization effects for the green and the black points. Deviations from $E^2/(p^2+m^2)$ are observed only for the coarsest ensembles
	   at high momenta, and even in the worst case they are well within the expected discretization errors, of order $O\left(\alpha_s\left(ap\right)^2\right)$.\label{2ptFigs}}
\end{figure}

As a consistency check, we test the continuum dispersion relation $E^2 = {\bf p}^2 + m ^2$ by plotting $E^2/({\bf p}^2 + m^2)$ vs ${\bf p}^2$ (fig.~\ref{2ptFigs}). Discretization effects
with large masses (in lattice units) could cause this quantity to deviate from $\sim 1$. Nonetheless, we clearly see an improvement as the lattice spacing is reduced.

\subsection{Three-point function ratios and general considerations}

We extract the matrix elements from ratios of 3-point correlation functions. To this end, we use the same setup and notation as in \cite{Bailey:2014tva,Lattice:2015rga}. We construct the
non-zero momentum correlation functions as
\begin{equation}
C^{\overline{B}\rightarrow D^\ast}_J({\bf p},t,T) = \left\langle\mathcal{O}_{D^\ast}({\bf -p},0)J({\bf p},t)\mathcal{O}_{B}({\bf 0},T)\right\rangle.
\end{equation}
where $\mathcal{O}_{D^\ast}({\bf p},0)$ and $\mathcal{O}_B({\bf 0},T)$ are interpolating operators with the quantum numbers of the $D^\ast$ and the $\overline{B}$ mesons, respectively,
with 3-momentum ${\bf p}$ and ${\bf 0}$, and $J$ is a vector or axial current. Then we can relate the correlation functions to the matrix elements using
\begin{equation}
C^{\overline{B}\rightarrow D^\ast}_J({\bf p},t,T) = \sqrt{Z_{D^\ast}({\bf p})Z_B({\bf 0})}e^{-E_{D^\ast}t}e^{-M_B\left(T-t\right)}
\left\langle D^\ast({\bf p})\big|J\big|B({\bf 0})\right\rangle + \ldots
\label{base3pt}
\end{equation}
The $Z_X$ factors are the overlap factors obtained from the 2-point functions of the meson $X$, and the exponentials decay with the energy of the constructed states. The missing
extra-terms refer to higher excited states, which are suppressed at large values of $t$ and $T - t$, as well as extra, oscillatory terms, whose contribution is proportional to $(-1)^t$
and $(-1)^{T-t}$. Regarding renormalization and matching, we follow references~\cite{Bailey:2014tva,Lattice:2015rga}, but at the time of the conference the perturbative $\rho$ factors
were not available yet.

In our calculations, the parent meson is always at rest and $\overline{B}$ propagates from time $T$. The daugther meson propagates from time $0$, and both of them are tied together at the
current insertion point $t$. We fix the sink and move the insertion in the range $[0,T]$, expecting the excited states in \eqref{base3pt} to die as we move far from the extremes. The
different source-sink separations used in this analysis are listed in table IV of \cite{Bailey:2014tva}.

From \eqref{base3pt} we construct ratios of 3-point functions that remove as many $Z$ factors and exponentials as possible from the rhs of \eqref{base3pt}. In order to suppress the effect
of the oscillatory terms coming from the 3-point functions, we use the following weighted average \cite{Bernard:2008dn}
\begin{equation}
\bar{R}(t,T) = \frac{1}{2}R(t,T) + \frac{1}{4} R(t,T+1) + \frac{1}{4} R(t+1,T+1),
\end{equation}
where $R$ is the ratio to be analyzed, $T$ is the sink time and $\bar{R}$ is the weighted average we use for the analysis. This average ensures that the errors introduced by oscillating
states are negligible compared with other errors. As a general ansatz to fit the ratios, we use the following function:
\begin{equation}
  r(t) = R\left(1 + A e^{-\Delta E_{M_0} t} + B e^{-\Delta E_{M_T} (T-t)}\right),
  \label{ansatz}
\end{equation}
where $R$ is the ratio of matrix elements we want to extract, $T$ is the sink time, $\Delta E_{M_0}$ and $\Delta E_{M_T}$ are the differences between the energies of the first excited
and the ground state for the mesons living at $t=0$ and $t=T$ respectively, taken from the 2-point fits, and $A$ and $B$ are fit parameters.

In our analysis we use constrained fits with priors. Whenever the parameters are largely unknown or known with poor precision, a loose prior is set just to guide the fitter. In
constrast, whenever the parameters involve quantities obtained in previous fits, the priors are set to the fit outcome and are counted as data points. To remove autocorrelations in
our data, we compute the covariance matrix given to the fitter after processing the data with jackknife.

\begin{figure}[h]
  \centering
  \sidecaption
  \includegraphics[width=0.40\linewidth,angle=0]{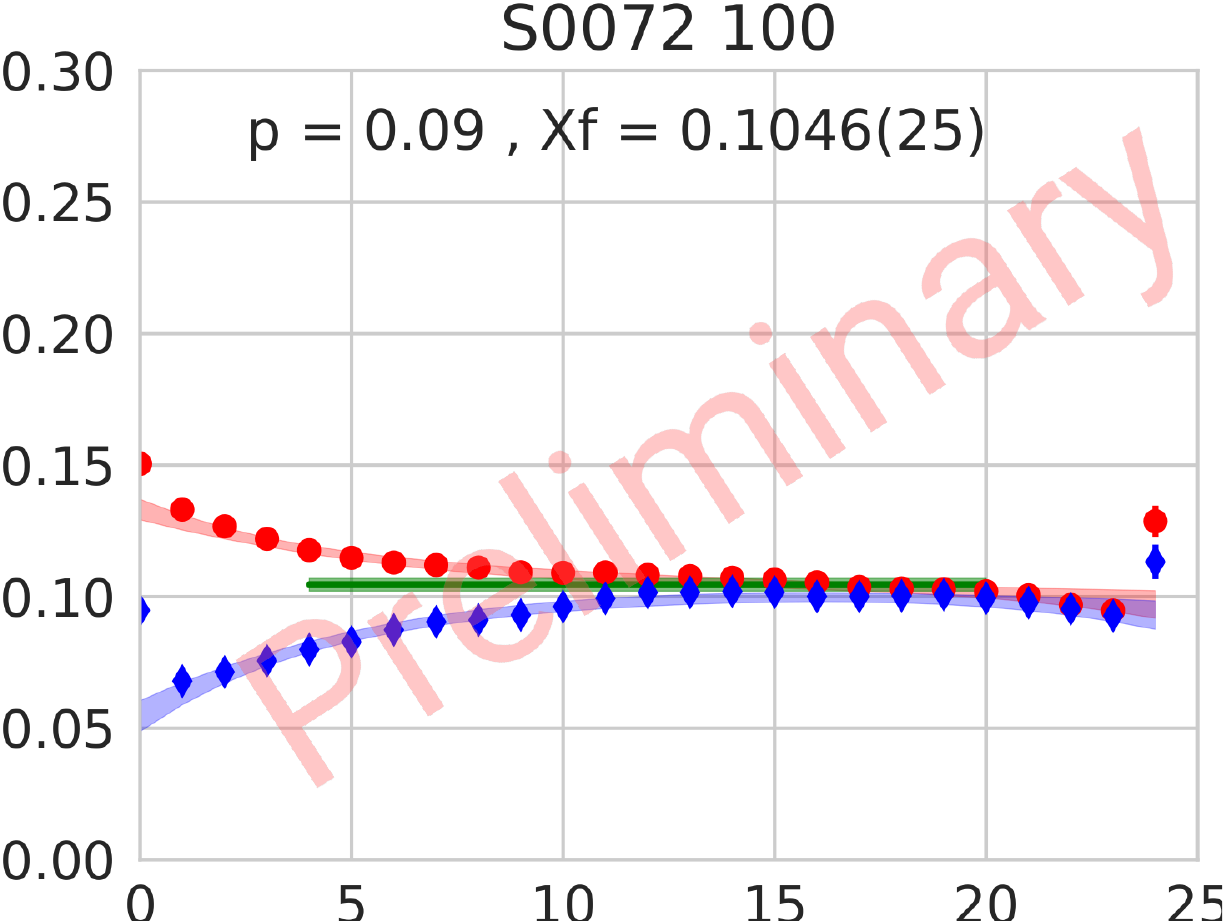}
  \caption{Sample ratio fit for a 3-point function ratio, $x_f$ in this case. The blue points show the data with a smeared interpolating operator, whereas the red points represent
  a point interpolating operator. The green horizontal bar represents the result for $R$ in \eqref{ansatz}. In this particular case, we force $B=0$ because the $\overline{B}$ smearing
  suppresses all the  excited states coming from the sink.\label{sFit}}
\end{figure}

Finally, the parent meson is always smeared using Richardson smearing, but the daughter meson can be smeared or not. This fact gives rise to several versions of each ratio, and
following the same technique as in the 2-point function analysis, we perform a joint fit of all the available data to obtain the final results. As an example, we show in fig.~\ref{sFit}
one of the fits we use to estimate the recoil parameter.
\section{Form factors from lattice matrix elements}

In order to extract the matrix elements from the lattice, we consider several ratios that reduce the errors in the determination of the form factors.

\subsection{Recoil parameter}\label{w3pt}
Considering the $\bar{B}$ meson at rest, one can compute the recoil parameter as

\smallskip

\begin{minipage}{0.42\linewidth}
  \begin{equation}
    w({\bf p}) = \frac{1+{\bf x}_f^2({\bf p})}{1-{\bf x}_f^2({\bf p})},
  \end{equation}
\end{minipage}
\hfill
\begin{minipage}{0.58\linewidth}
  \begin{equation}
    {\bf x}_f({\bf p}) = \frac{\left\langle D^\ast({\bf p})\right|{\bf V}\left|D^\ast({\bf 0})\right\rangle}{\left\langle D^\ast({\bf p})\right|V^4\left|D^\ast({\bf 0})\right\rangle},
    \label{xfRatio}
  \end{equation}
\end{minipage}

\smallskip

\noindent where the ratio for ${\bf x}_f$ comes from the flavor-diagonal transition $D^\ast(p,s) \rightarrow D^\ast(p',s')$.

\subsection{Axial form factors}\label{axialSec}
\subsubsection{Axial double ratio $R_{A_1}$}
Although $R_{A_1}$ is directly obtained from $\left\langle D^\ast({\bf p}_\bot)\big|A_1\big|B(0)\right\rangle$, it is advantageous to use a ratio,
\begin{equation}
R_{A_1}({\bf p}) = \frac{\left\langle D^\ast({\bf p}_\bot)\big|A_1\big|\bar{B}(0)\right\rangle}{\left\langle D^\ast(0)\big|A_1\big|\bar{B}(0)\right\rangle}.
\label{sRat}
\end{equation}
Eq.~\eqref{sRat} has an important drawback: when expressed as a ratio of 3-point functions, exponentials depending on time are not cancelled for the daughter meson. Hence, we need to
add an energy-dependent correction factor before we can fit the ratio on the lattice. In the zero momentum case a double ratio was used to get rid of the exponential and some overlap
factors, reducing the errors and improving the quality of the fit \cite{Bernard:2008dn,Bailey:2014tva}. Here we can do the same,
\begin{equation}
\left|R_{A_1}({\bf p})\right|^2 = \frac{\left\langle D^\ast({\bf p}_\bot)\big|A_1\big|B(0)\right\rangle\left\langle B(0)\big|A_1\big|D^\ast({\bf p}_\bot)\right\rangle}
{\left\langle D^\ast(0)\big|V_4\big|B(0)\right\rangle\left\langle B(0)\big|V_4\big|D^\ast(0)\right\rangle}.
\label{dRat}
\end{equation}
From this quantity we can directly extract $h_{A_1}(w) = 2/(1+w)R_{A_1}$. As we don't have any measurements for the matrix element
$\left\langle B(0)\big|A_1\big|D^\ast({\bf p}_\bot)\right\rangle$, we use time reversal $\mathcal{T}$ to reconstruct it from known data,
\begin{equation}
C^{D^\ast\rightarrow\overline{B}}({\bf p}_\bot, t,T) \xrightarrow{\,\mathcal{T}\,} C^{\overline{B}\rightarrow D^\ast}_{A_1}({\bf p}_\bot, T-t,T).
\end{equation}
Our preliminary results for $h_{A_1}(w)$, computed from eq.~\eqref{dRat} is shown in the right pane of fig.~\ref{allFf1}.

\subsubsection{Ratios $R_0$ and $R_1$}
The quantities $R_0$ and $R_1$

\begin{minipage}{0.481\textwidth}
  \begin{equation}
    R_0({\bf p}) = \frac{\left\langle D^\ast({\bf p}_\parallel)\big|A_4\big|B(0)\right\rangle}{\left\langle D^\ast({\bf p}_\bot)\big|A_1\big|B(0)\right\rangle},
  \end{equation}
\end{minipage}%
\begin{minipage}{0.481\textwidth}
  \begin{equation}
    R_1({\bf p}) = \frac{\left\langle D^\ast({\bf p}_\parallel)\big|A_1\big|B(0)\right\rangle}{\left\langle D^\ast({\bf p}_\bot)\big|A_1\big|B(0)\right\rangle},
  \end{equation}
\end{minipage}

\noindent encode the behavior of $h_{A_2}(w)$ and $h_{A_3}(w)$ as

\begin{minipage}{0.481\textwidth}
  \begin{equation}
    R_0 = \frac{\sqrt{w^2-1}(1-h_{A_2}+wh_{A_3})}{(1+w)h_{A_1}},
  \end{equation}
\end{minipage}%
\begin{minipage}{0.481\textwidth}
  \begin{equation}
    R_1 = w-\frac{(w^2-1)h_{A_3}}{(1+w)h_{A_1}}.
  \end{equation}
\end{minipage}
Preliminary results for $h_{A_2}(w)$ and $h_{A_3}(w)$ are reported in fig.~\ref{allFf2}.
%\begin{minipage}{0.481\textwidth}
%  \begin{equation}
%    R_0^{s}(t) = \sqrt{\frac{Z_{D^\ast}(p_\parallel,s)}{Z_{D^\ast}(p_\bot,s)}}\frac{\left\langle D^\ast(p_\parallel)\big|A_4\big|B(0)\right\rangle}{\left\langle D^\ast(p_\bot)\big|A_1\big|B(0)\right\rangle}, \\
%  \end{equation}
%\end{minipage}%
%\begin{minipage}{0.481\textwidth}
%  \begin{equation}
%    R_1^{s}(t) = \sqrt{\frac{Z_{D^\ast}(p_\parallel,s)}{Z_{D^\ast}(p_\bot,s)}}\frac{\left\langle D^\ast(p_\parallel)\big|A_1\big|B(0)\right\rangle}{\left\langle D^\ast(p_\bot)\big|A_1\big|B(0)\right\rangle},
%  \end{equation}
%\end{minipage}

%\smallskip

\subsection{Vector form factor}\label{vectorSec}
The previously defined quantity $X_V(w)$ can be measured as the following ratio of matrix elements:

\begin{minipage}{0.42\linewidth}
  \begin{equation}
    h_V = \frac{R_{A_1}}{\sqrt{w-1}}X_V
  \end{equation}
\end{minipage}
\begin{minipage}{0.58\linewidth}
  \begin{equation}
    X_V({\bf p}) = \frac{\left\langle D^\ast({\bf p}_\bot)\big|V_1\big|B(0)\right\rangle}{\left\langle D^\ast({\bf p}_\bot)\big|A_1\big|B(0)\right\rangle}.
  \end{equation}
\end{minipage}
When this ratio is expressed in terms of lattice 3-point correlation functions, all the exponentials and overlap factors are cancelled, and the result yields directly the quotient of
form factors we are looking for. Our preliminary result for $h_V(w)$ is shown in the left pane of fig.~\ref{allFf1}.

\subsection{Results for the form factors as a function of the recoil parameter}\label{ffRes}
\begin{figure}[h]
  \centering
  \subfigure[$h_V(w)$ form factor.]
            {\includegraphics[width=0.48\linewidth,angle=0]{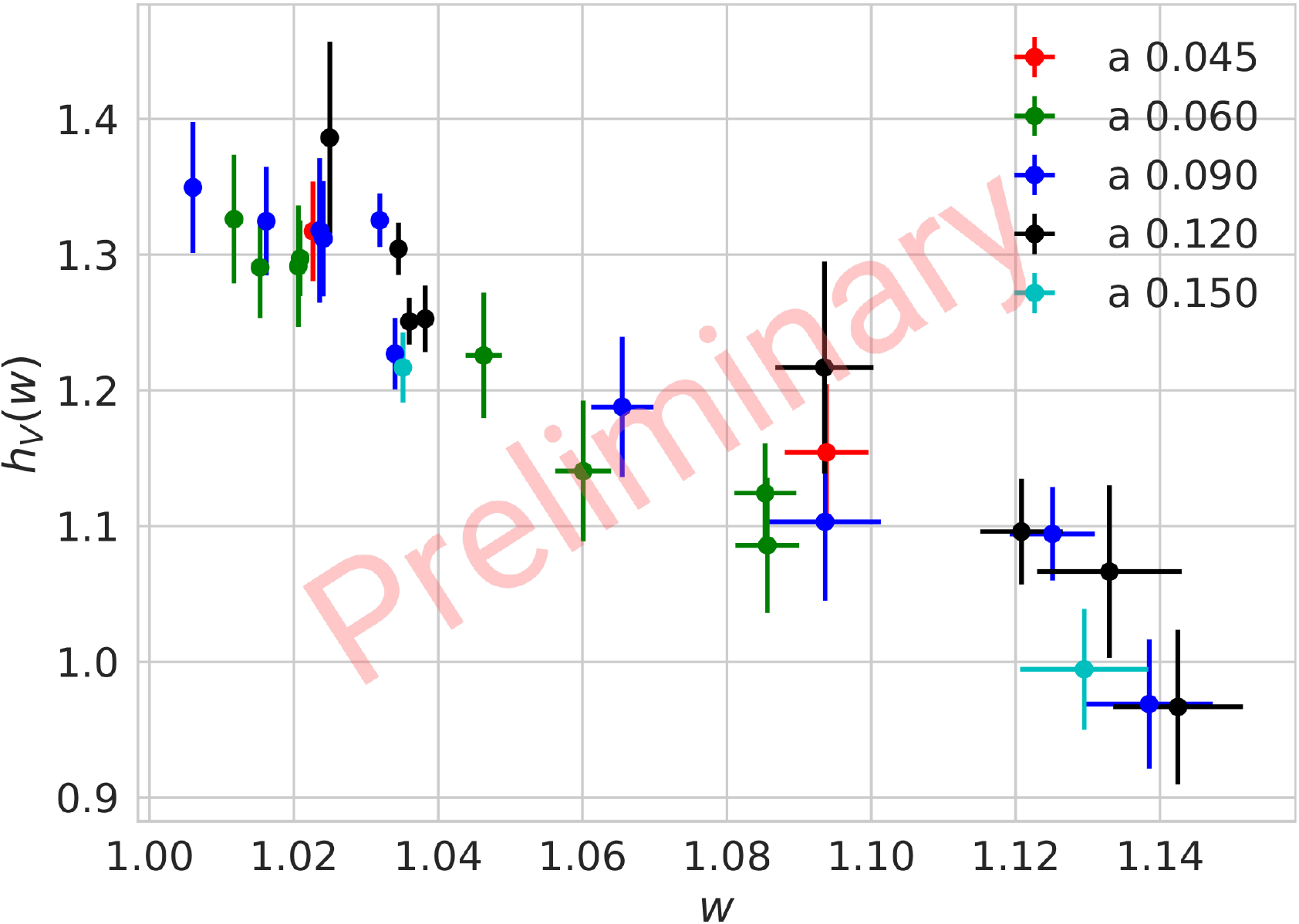} }
  \subfigure[$h_{A_1}(w)$ form factor.]
            {\includegraphics[width=0.48\linewidth,angle=0]{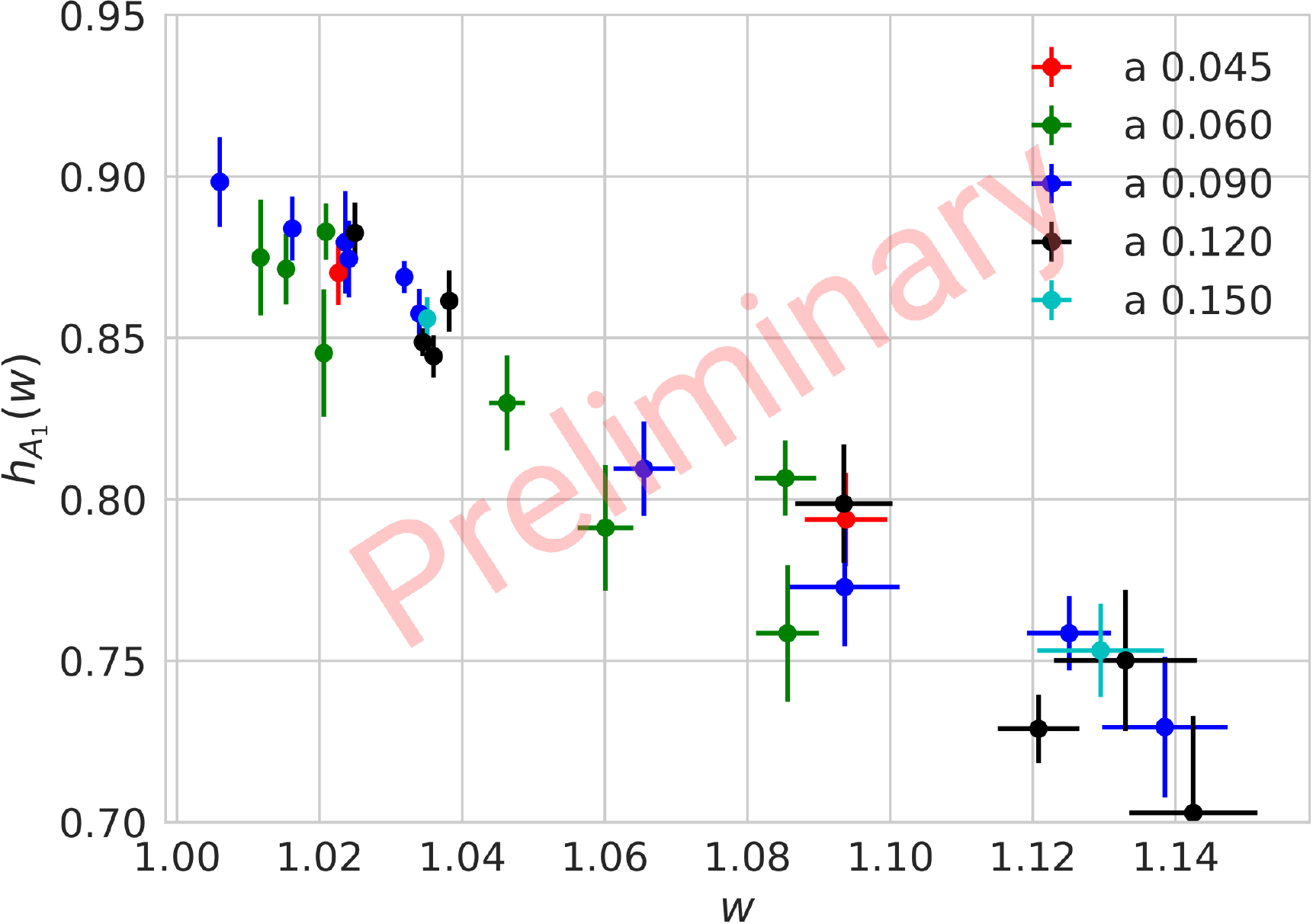} }
  \caption{Preliminary results for $h_V(w)$ and $h_{A_1}(w)$ from $X_V$ and the double ratio $R_{A_1}$. The contribution of $h_V(w)$ to $\mathcal{F}(w)$ is suppressed due to kinematic
	   factors, and the final result is clearly dominated by $h_{A_1}(w)$.\label{allFf1}}
\end{figure}

\begin{figure}[h]
  \centering
  \subfigure[$h_{A_2}(w)$ form factor.]
            {\includegraphics[width=0.48\linewidth,angle=0]{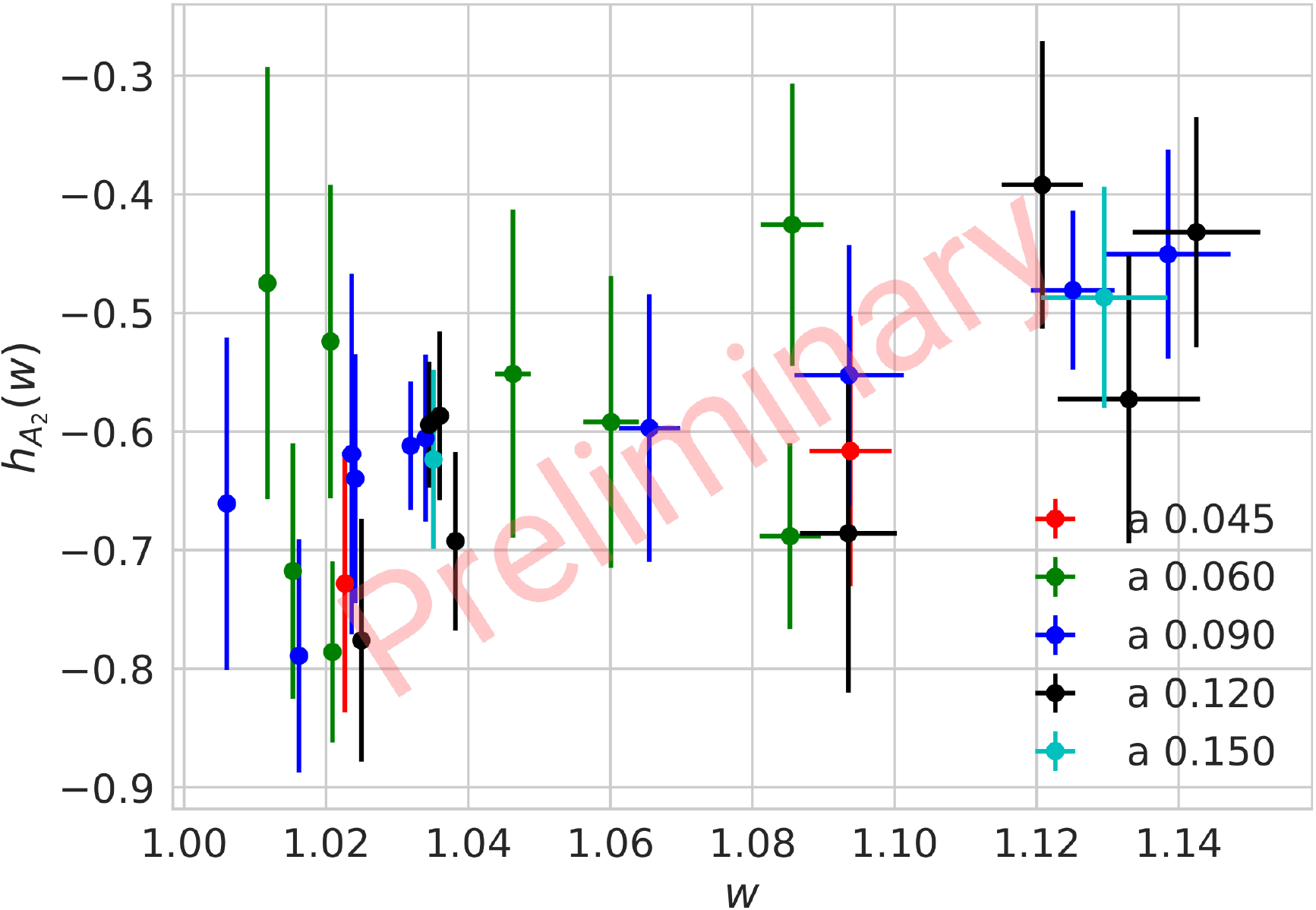} }
  \subfigure[$h_{A_3}(w)$ form factor.]
            {\includegraphics[width=0.48\linewidth,angle=0]{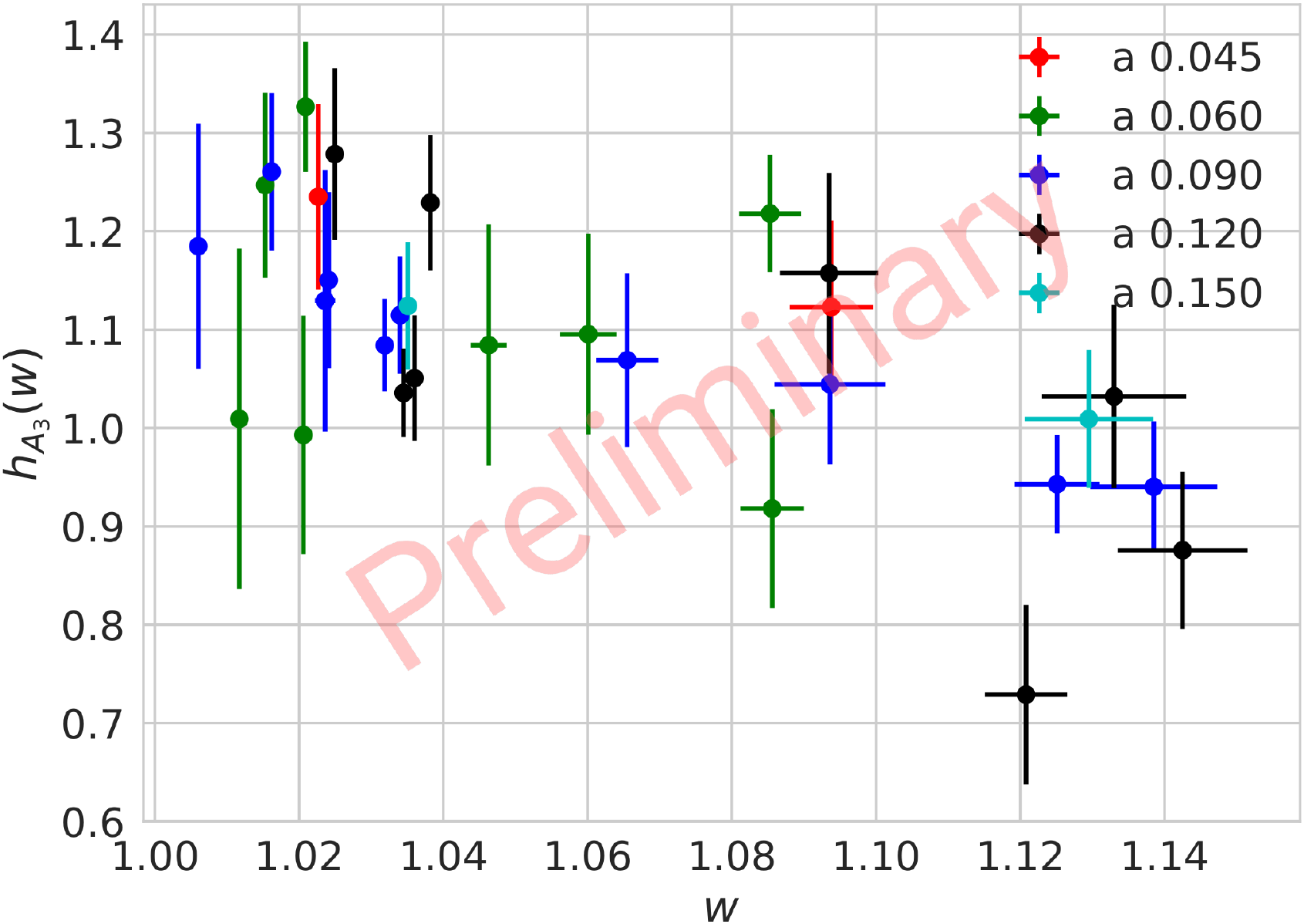} }
  \caption{Preliminary results for $h_{A_2}(w)$ and $h_{A_3}(w)$ from $R_0$ and $R_1$. We expect to reduce the errors in these form factors when we refine the analysis. However, the
	   contribution of $h_{A_2}(w)$ and $h_{A_3}(w)$ to $\mathcal{F}(w)$ is highly suppressed by kinematic factors at low recoil. Therefore the final total error is dominated by
	   the errors coming from $h_{A_1}(w)$.\label{allFf2}}
\end{figure}

In figs.~\ref{allFf1},\ref{allFf2} we show the preliminary results for the axial and vector form factors, with statistical errors only, without rho factors, and before taking the
chiral-continuum extrapolation. It is important to notice that near zero recoil (for small $w-1$), $\mathcal{F}(w)$ is dominated by $h_{A_1}$, because the contributions from $h_V$,
$h_{A_2}$, and $h_{A_3}$ are suppressed by kinematic factors (see eqs.~\eqref{X2},~\eqref{X3} and~\eqref{XV}).

\section{Summary and future work}\label{summ}

In this paper we have presented first preliminary results from our lattice QCD calculations of the form factors for $\overline{B}\rightarrow D^\ast\ell\overline{\nu}$ at non-zero recoil.
Still to be completed are further improvements in the fits to the correlation functions, after which we plan to study the chiral-continuum extrapolations, use the $z$-expansion to
parametrize the shape, and construct a complete, systematic error budget.

\bigskip

\begin{acknowledgement}
\section{Acknowledgements}\label{ack}

Computations for this work were carried out with resources provided by the USQCD Collaboration, the National Energy Research Scientific Computing Center and the Argonne Leadership
Computing Facility, which are funded by the Office of Science of the U.S. Department of Energy; and with resources provided by the National Institute for Computational Science and the
Texas Advanced Computing Center, which are funded through the National Science Foundation’s Teragrid/XSEDE Program. This work was supported in part by the U.S. Department of Energy under
grants No. DE-FC02-06ER41446 (C.D.) and No. DE-SC0015655 (A.X.K.), by the U.S. National Science Foundation under grants PHY10-67881 and PHY14-17805 (J.L.), PHY14-14614 (C.D., A.V.); by
the Fermilab Distinguished Scholars program (A.X.K.); by the German Excellence Initiative and the European Union Seventh Framework Program under grant agreement No. 291763 as well as the
European Union’s Marie Curie COFUND program (A.S.K.).

Fermilab is operated by Fermi Research Alliance, LLC, under Contract No. DE-AC02-07CH11359 with the United States Department of Energy, Office of Science, Office of High Energy Physics.
The publisher, by accepting the article for publication, acknowledges that the United States Government retains a non-exclusive, paid-up, irrevocable, world-wide license to publish or
reproduce the published form of this manuscript, or allow others to do so, for United States Government purposes.

\end{acknowledgement}

\bibliography{lattice2017}

\begin{thebibliography}{13}

\bibitem{Cabibbo:1963yz}
N.~Cabibbo, Phys. Rev. Lett. \textbf{10}, 531 (1963), [,648(1963)]

\bibitem{Kobayashi:1973fv}
M.~Kobayashi, T.~Maskawa, Prog. Theor. Phys. \textbf{49}, 652 (1973)

\bibitem{Antonelli:2009ws}
M.~Antonelli et~al., Phys. Rept. \textbf{494}, 197 (2010), \texttt{0907.5386}

\bibitem{Patrignani:2016xqp}
C.~Patrignani et~al. (Particle Data Group), Chin. Phys. \textbf{C40}, 100001
  (2016)

\bibitem{Amhis:2016xyh}
Y.~Amhis et~al. (2016), \texttt{1612.07233}

\bibitem{Aaij:2244311}
R.~Aaij, B.~Adeva, M.~Adinolfi, Z.~Ajaltouni, S.~Akar, J.~Albrecht, F.~Alessio,
  M.~Alexander, S.~Ali, G.~Alkhazov et~al. (LHCb Collaboration), Tech. Rep.
  CERN-LHCC-2017-003, CERN, Geneva (2017),
  \urlstyle{tt}\url{http://cds.cern.ch/record/2244311}

\bibitem{Abe:2010gxa}
T.~Abe et~al. (Belle-II), Tech. rep., KEK (2010), \texttt{1011.0352},
  \urlstyle{tt}\url{http://arxiv.org/pdf/1011.0352v1}

\bibitem{Bernard:2008dn}
C.~Bernard et~al., Phys. Rev. \textbf{D79}, 014506 (2009), \texttt{0808.2519}

\bibitem{Bailey:2014tva}
J.A. Bailey et~al. (Fermilab Lattice, MILC), Phys. Rev. \textbf{D89}, 114504
  (2014), \texttt{1403.0635}

\bibitem{Bazavov:2009bb}
A.~Bazavov et~al. (MILC), Rev. Mod. Phys. \textbf{82}, 1349 (2010),
  \texttt{0903.3598}

\bibitem{Lepage:1998vj}
G.P. Lepage, Phys. Rev. \textbf{D59}, 074502 (1999), \texttt{hep-lat/9809157}

\bibitem{ElKhadra:1996mp}
A.X. El-Khadra, A.S. Kronfeld, P.B. Mackenzie, Phys. Rev. \textbf{D55}, 3933
  (1997), \texttt{hep-lat/9604004}

\bibitem{Lattice:2015rga}
J.A. Bailey et~al. (MILC), Phys. Rev. \textbf{D92}, 034506 (2015),
  \texttt{1503.07237}

\end{thebibliography}

\end{document}